\documentclass[dvips,graphicx,11pt]{article}
\usepackage{lscape,graphicx}
\usepackage{color}
\usepackage{amssymb}
\usepackage{euler}

\usepackage{mathptmx}      
\usepackage{natbib}  

\newcommand{\aap}{{Astron. Astrophys.}}
\newcommand{\apj}{{Astrophys. J.}}
\newcommand{\jgr}{{J. Geophys. Res.}}

\newcommand{\solphys}{{Solar Phys.}}

\usepackage{natbib}
\usepackage{pifont}
\usepackage{amsmath}

\begin{document}

\begin{center}
{\large\bf 
{ Phase mixing of  Alfv\'en waves near a 2D magnetic null point} }

\bigskip
{\large\bf {J.~A.~{McLaughlin} } }

\bigskip
{Department of Mathematics \& Information Sciences, Northumbria University, Newcastle Upon Tyne, NE1 8ST, UK}
\end{center}

\bigskip
{\bf Abstract:}

{

The propagation of linear Alfv\'en wave pulses in an inhomogeneous plasma    near a 2D coronal null point is investigated.  When a uniform plasma     density is considered, it is seen that an initially planar Alfv\'en wavefront     remains planar, despite the varying equilibrium Alfv\'en speed, and  that  all the wave collects    at the separatrices. Thus, in the non-ideal case, these Alfv\'enic disturbances   preferentially   dissipate    their energy at these locations. For a  non-uniform  equilibrium density, it is found that the Alfv\'en wavefront is significantly distorted away from the initially planar geometry, inviting the possibility of dissipation due to phase mixing. Despite this however,  we conclude that for the Alfv\'en wave, current density accumulation and preferential heating still primarily occur at the separatrices, {{even when an extremely non-uniform density profile is considered}}.

}




\section{Introduction}

MHD wave propagation within an {{ inhomogeneous medium }} is a fundamental plasma process and the study of MHD waves in the neighbourhood of magnetic null points directly contributes to this area (see McLaughlin {\emph{et al.}} \citeyear{McLaughlinREVIEW} for a comprehensive review of the topic). {\emph{Null points}} are  weaknesses in the magnetic field at which  the field strength, and thus the Alfv\'en speed, is zero. {\emph{Separatrices}} are topological features that separate regions of different magnetic connectivity, and are an inevitable consequence of the isolated magnetic flux fragments  in the photosphere. The number of resultant null points  depends upon the complexity of the magnetic flux distribution, but tens of thousands are estimated to be present (see, e.g., Close {\emph{et al.}} \citeyear{Close2004}; Longcope \citeyear{L2005}; R{\'e}gnier {\emph{et al.}} \citeyear{RPH2008}; Longcope \& Parnell \citeyear{LP2009}).

\vspace{0.05cm}

In addition, MHD wave perturbations are ubiquitary in the solar corona (e.g. Tomczyk {\emph{et al.}} \citeyear{Tomczyk2007}) and a variety of  observations have clearly demonstrated the existence of wave activity for all three of the basic waves modes; namely Alfv\'en waves and fast and slow magnetoacoustic waves. That the waves exist is no longer in doubt but the surprising fact is that these are generally rapidly damped. Waves in a uniform magnetic field and plasma have extremely long damping lengths and so the explanation for the observations must lie  in the non-uniform nature of the solar corona.{{ Non-thermal line broadening and narrowing  due to  Alfv\'en waves has been reported by various authors, including Banerjee {\emph{et al.}} (\citeyear{Banerjee1998}), Erd\'elyi  {\emph{et al.}} (\citeyear{Erdelyi1998}),  Harrison {\emph{et al.}} (\citeyear{Harrison2002}) and  O'Shea {\emph{et al.}} (\citeyear{OShea2003}; \citeyear{OShea2005}). The role of  Alfv\'en waves in coronal heating through dissipation and observed spectral line broadening has been reported both analytically (e.g. Dwivedi  \& Srivastava \citeyear{DS2006}) and more recently numerically (e.g. Chmielewski  {\emph{et al.}} {\citeyear{Chmielewski2013}}, and references therein).

}}

\vspace{0.05cm}

Thus, MHD waves and magnetic topology {\emph{will}} encounter each other in the corona (e.g. waves emanating from a flare or CME will at some point encounter a coronal null point). The behaviour of linear MHD waves (fast \& slow magnetoacoustic waves and Alfv\'en waves) has been investigated in the neighbourhood of a variety of 2D null points (e.g. McLaughlin \& Hood \citeyear{MH2004}; \citeyear{MH2005}; \citeyear{MH2006a}; \citeyear{MH2006b}; McLaughlin {\emph{et al.}} \citeyear{MFH2008}). These authors   found that the (linear) Alfv\'en wave propagates along magnetic fieldlines and accumulates along the separatrices in 2D, or along the spine or fan-plane in 3D. Thus, these authors make a key prediction: {\emph{separatrices, spines and/or fan-planes will be locations for preferential heating by (linear) Alfv\'en waves}}. 

\vspace{0.05cm}

Waves in the neighbourhood of a single 2D null point have also been investigated using cylindrical models, in which  the generated waves encircled the null point (e.g.  Bulanov \& Syrovatskii \citeyear{Bulanov1980};   Craig \& McClymont \citeyear{CM1991}; \citeyear{CM1993}; Craig \& Watson \citeyear{CW1992}; Hassam \citeyear{Hassam1992}) and it was found that the wave propagation leads to an  exponentially-large increase in the current density (see also Ofman {\emph{et al.}} \citeyear{Ofman1993}; Steinolfson {\emph{et al.}} \citeyear{Steinolfson1995} and a comprehensive review by  McLaughlin {\emph{et al.}} \citeyear{McLaughlinREVIEW} for further details). Nonlinear and three-dimensional MHD wave activity about coronal null points has also been investigated (e.g. Galsgaard {\emph{et al.}} \citeyear{Galsgaard2003};   Pontin \& {Galsgaard} \citeyear{PG2007};  Pontin {\emph{et al.}} \citeyear{PBG2007}; McLaughlin {\emph{et al.}} \citeyear{MFH2008};  \citeyear{McLaughlin2009}; Galsgaard \& Pontin \citeyear{klaus2011a}; \citeyear{klaus2011b}; Thurgood \& McLaughlin \citeyear{Thurgood2012}; \citeyear{Thurgood2013}).

\vspace{0.05cm}


One of the most efficient damping   {{ mechanisms of Alfv\'en waves }}   to date is called {\emph{phase mixing}} and is described by Heyvaerts \& Priest (\citeyear{HP1983}) for a harmonic wave train propagating in a uniform vertical magnetic field. They found that the amplitudes decay as the negative exponential depending on the third power of the height and linearly with magnetic resistivity, $\eta$. Thus, the damping length depends on $\eta^{-1/3}$. Since observations rarely show more than a few periods at a time, Hood {\it et al.} (\citeyear{Hood2002}) investigated the propagation of single pulses and found that the decay was now algebraic in nature but still dependent on $\eta^{-1/3}$.

\vspace{0.05cm}

The phase mixing mechanism is simple to explain; when the plasma has a density gradient perpendicular to the magnetic field, the Alfv\'en speed is a function of the transverse coordinate. Thus, the Alfv\'en waves propagate on each fieldline with their own local Alfv\'en speed. After a certain time, the Alfv\'en wave perturbations on neighbouring fieldlines become out of phase (e.g. Botha {\emph{et al.}} \citeyear{Botha2000}; McLaughlin {\emph{et al.}} \citeyear{McLaughlin2011b}). It is precisely the Alfv\'en perturbations oscillating independently from their neighbours that leads to the build-up of small length scales and consequently current generation and hence dissipation.

\vspace{0.05cm}

In this paper, we will investigate the behaviour of the linear Alfv\'en wave in the neighbourhood of a simple 2D X-point geometry, and we shall consider the behaviour in both uniform and non-uniform density plasma. This lifts one of the key restrictions imposed by McLaughlin \& Hood (\citeyear{MH2004}) and its subsequent papers, namely the assumption of constant equilibrium density.  With a non-uniform density profile,  the Alfv\'en speed is now changing from fieldline to fieldline, and thus we may have phase mixing. This is the key question that this paper addresses: with the addition of a non-uniform density,  does the current build-up still occur at the separatrix or does phase mixing now allow the energy to be extracted from a different location?

\vspace{0.05cm}

Our paper has the following outline: the basic setup, equations and assumptions are described in \S\ref{governing_equations}, the numerical and analytical results are presented in \S\ref{sec:6.4.1}, and the conclusions are given in \S\ref{sec:6.10}.


\section{Basic Equations}\label{governing_equations}

The MHD equations for a low-$\beta$ plasma appropriate to the solar corona are used. Hence,
\begin{eqnarray}
\rho \left[ {\partial {\bf{v}}\over \partial t} + \left( {\bf{v}}\cdot\nabla \right) {\bf{v}} \right] &=& {1\over \mu}\left(\nabla \times {\bf{B}}\right)\times {\bf{B}},\label{eq:2.1} \\
  {\partial {\bf{B}}\over \partial t} &=& \nabla \times \left
  ({\bf{v}}\times {\bf{B}}\right ) + \eta \nabla^2
  {\bf{B}},\nonumber \\
  {\partial \rho\over \partial t} + \nabla \cdot \left (\rho {\bf{v}}\right )
  &=& 0, \qquad \nabla \cdot {\bf{B}} =0 ,\nonumber
\end{eqnarray}
where $\rho$ is the mass density, ${\bf{v}}$ is the plasma velocity, ${\bf{B}}$ the magnetic induction (usually called the magnetic field), $ \mu = 4 \pi \times 10^{-7} \/\mathrm{Hm^{-1}}$  the magnetic permeability, $\eta = 1/\mu\sigma$ is the magnetic diffusivity $\left( \mathrm{m}^2\mathrm{s}^{-1}\right)$, and $\sigma$ the electrical conductivity. The gas pressure and the adiabatic energy equation are neglected in the low-$\beta$ approximation. {{  We have also neglected viscous terms in equations (\ref{eq:2.1}). Investigations involving viscous magnetofluids can be found in  Kumar \& Bhattacharyya  (\citeyear{KB2011}) and McLaughlin {\emph{et al.}} (\citeyear{McLaughlin2011b}) and references therein.

}}

\vspace{0.05cm}

\begin{figure}
\begin{center}
\includegraphics[width=2.2in]{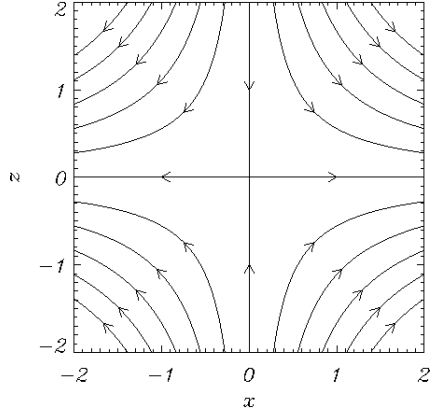}
\caption{Equilibrium magnetic field.}
\label{figure1}
\end{center}
\end{figure}

The equilibrium  magnetic field structure is taken as a simple 2D X-type neutral point:
\begin{equation}\label{eq:2.2}
\qquad {\bf{B}}_0 = B_0 \left({x\over L}, 0, -{z\over L}\right),
\end{equation}
where $B_0$ is a characteristic field strength and $L$ is the length scale for magnetic field variations. This magnetic field can be seen in Figure \ref{figure1}.  Note that this particular choice of magnetic field is only valid in the neighbourhood of the null point located at $x=z=0$.{{  In addition, equation (\ref{eq:2.2}) is potential, although in general coronal fields are  twisted and thus a potential field is a coarse approximation. }}

We can also write ${\bf{B}}_0 = \nabla \times {\bf{A}}$ where $ {\bf{A}} = (0,A_0,0)$ is the {{vector potential}}. For our particular choice of equilibrium magnetic field,  $A_0=-xz$.

\subsection{Linearised equations}\label{sec:2.2}

To study the nature of wave propagation near null points, the linearised MHD equations are used. Using subscripts of $0$ for equilibrium quantities and $1$ for perturbed quantities, the linearised versions of equations (\ref{eq:2.1}) are:
\begin{eqnarray}
\rho_0 \frac{\partial \mathbf{v}_1}{\partial t} &=& {1 \over \mu} \left( \nabla \times {\mathbf{B}_1} \right) \times \mathbf{B}_0 \; ,\label{eq:2.3} \\
{\partial {\bf{B}}_1\over \partial t} &=& \nabla \times    ({\bf{v}}_1 \times {\bf{B}}_0) + \eta \nabla^2 {\bf{B}}_1 \; ,\nonumber \\
\frac{\partial \rho_1} {\partial t} + \nabla\cdot\left( \rho_0 \mathbf{v} _1 \right) &=& 0 \; , \qquad \nabla \cdot {\bf{B}}_1 =0 \;. \nonumber
\end{eqnarray}
We will not discuss the linearised continuity equation further as it can be solved once we know $\mathbf{v} _1$. In fact, it has no influence on the momentum equation (in the low $\beta$ approximation) and so in effect the plasma is arbitrarily compressible (e.g. Craig \& Watson \citeyear{CW1992}).

\vspace{0.05cm}

We now consider a change of scale to non-dimensionalise; let ${\mathbf{\mathrm{v}}}_1 = \bar{\rm{v}} {\mathbf{v}}_1^*$, ${\mathbf{B}}_0 = B_0 {\mathbf{B}}_0^*$, ${\mathbf{B}}_1 = B_0 {\mathbf{B}}_1^*$, $x = L\: x^*$, $z=Lz^*$, $\nabla = \frac{1}{L}\nabla^*$ and $t=\bar{t}t^*$, where we let * denote a dimensionless quantity and $\bar{\rm{v}}$, $B_0$, $L$ and $\bar{t}$ are constants with the dimensions of the variable they are scaling. We then set $ {B_0} / {\sqrt{\mu \rho _0 } } =\bar{\rm{v}}$ and $\bar{\rm{v}} =  {a} / {\bar{t}}$ (this sets $\bar{\rm{v}}$ as the background Alfv\'{e}n speed). This process non-dimensionalises equations (\ref{eq:2.3})  and under these scalings, $t^*=1$ (for example) refers to $t=\bar{t}=  {L} / {\bar{\rm{v}}}$; i.e. the (background) Alfv\'en time taken to travel a distance $L$. For the rest of this paper, we drop the star indices; the fact that they are now non-dimensionalised is understood.


\vspace{0.05cm}

We now restrict our attention to 2.5D MHD, i.e. 3D MHD with an invariant direction, and here we arbitrarily take $\partial / \partial y =0$. {{ In addition, from now on we consider an ideal plasma (i.e. let $\eta=0$ or $R_m \rightarrow \infty$) but will discuss the role of resistivity further in the conclusions. Numerical diffusion, although present in all numerical simulations, plays a negligible role.}} The linearised MHD equations (\ref{eq:2.3}) naturally decouple into two sets of equations, with one set  governing the behaviour in the invariant direction (i.e. here the $y-$direction) and the other governing behaviour in the $xz-$plane only. Furthermore, McLaughlin \& Hood (\citeyear{MH2004}) showed that the behaviour in the invariant direction corresponded to Alfv\'en wave behaviour, and that the equations in the $xz-$plane governed the fast MHD wave behaviour (note the slow MHD wave is absent in the low-$\beta$ limit).

\vspace{0.05cm}

In this paper, we focus on the linearised equations for the Alfv\'en wave, with ${\mathbf{v}} _1 = \left( v_x, v_y, v_z \right)$ and ${\mathbf{B}} _1 = \left( b_x, b_y, b_z \right)$. For details of the fast wave equations see the review by McLaughlin {\emph{et al.}} (\citeyear{McLaughlinREVIEW}).

\vspace{0.05cm}

The equations governing the behaviour in the invariant direction (i.e. the $y-$direction) are:
\begin{eqnarray}
\rho_0 \frac{\partial {{{v}_y}}}{\partial t} &=& \left( { \mathbf{B}}_0 \cdot \nabla \right)  b_y  =   \left(B_x \frac {\partial }{\partial x} +B_z\frac {\partial }{\partial z} \right) b_y             \;,\nonumber\\
\frac{\partial {b}_y }{\partial t} &=&  \left( { \mathbf{B}}_0 \cdot \nabla \right) v_y + \frac{1}{R_m}\nabla^2 {{b}}_y =  \left(B_x \frac {\partial }{\partial x} + B_z \frac {\partial }{\partial z} \right) v_y    + \frac{1}{R_m}  \left( \frac{\partial^2}{\partial x^2 }+  \frac{\partial^2}{\partial z^2 }\right)   b_y          \label{xmen}  \;,
\end{eqnarray}
where $v_y$ is the velocity out of the plane that ${\bf{B}}_0$ defines. Hence, waves with this velocity will be {\emph{transverse}} waves (energy flow perpendicular to the wavevector).


\vspace{0.05cm}

We will now vary the background plasma density. A straightforward  way to add a non-uniform density  profile to the governing equations is to consider $\rho_0=\rho_0(A_0)$, where ${\bf{A}}=(0,A_0,0)$ is the {{vector potential}} and $A_0$ is its $y-$coordinate. Thus, since $\nabla A_0$ is perpendicular to ${\bf{B}}_0$, we have a density gradient  perpendicular to the magnetic field. Under this model, $\rho_0$ is now constant {\emph{along}} a fieldline but it can vary {\emph{across}} fieldlines, i.e. vary from fieldline to fieldline. Hence, the {{equilibrium}} density is purely a function of $A_0$, namely
\begin{equation}\label{eq:density_PM}
\rho_0 = \rho_0(A_0) = \rho_0(xz)\;,
\end{equation}
where $A_0=-xz$ {{(recall that at this point all our variables are non-dimensionalised)}}. Thus equations (\ref{xmen}) can be combined and written as: 
\begin{eqnarray}
  \frac {\partial^2 v_y }{\partial t^2}  =  \frac{1}{\rho_0(xz)}  \left( { \mathbf{B}}_0 \cdot \nabla \right)^2 v_y                       =\frac{1}{\rho_0(xz)}  \left(B_x \frac {\partial }{\partial x} + B_z\frac {\partial }{\partial z} \right) ^2 v_y  =\frac{1}{\rho_0(xz)}  \left(x \frac {\partial }{\partial x} -z \frac {\partial }{\partial z} \right) ^2 v_y \; \label{alfvenalpha_PM},
\end{eqnarray}
where we have taken $\eta=0$ and have implemented our choice of ${\bf{B}}_0$ from equation (\ref{eq:2.2}).

\vspace{0.05cm}

This is the primary equation we will be utilising in this paper to investigate the behaviour of the Alfv\'en wave. In its derivation, we have assumed linearised behaviour in an ideal  2.5D plasma, i.e. a 3D plasma with an invariant direction.

\vspace{0.05cm}

If we now define $V_{A0}(xz) = {1} / {\sqrt{\rho_0(xz)}}$  then equation (\ref{alfvenalpha_PM}) can be written as:
\begin{eqnarray}
  \frac {\partial^2 v_y }{\partial t^2} =  V_{A0}(xz)^2   \left(x \frac {\partial }{\partial x} - z\frac {\partial }{\partial z} \right)  ^2 v_y                    =  V_{A0}(xz)^2     \left( { \mathbf{B}}_0 \cdot \nabla \right)^2 v_y                   \label{equation_alfven_wave}
\end{eqnarray}
Here  $V_{A0}(xz)$ is related to the equilibrium (Alfv\'en) speed of the system. It is precisely this non-constant Alfv\'en speed, i.e. $V_{A0} =V_{A0} \: (xz)$, that leads to gradients in the  Alfv\'en-speed profile, and hence to the possibility of phase mixing. Note that $\rho_0$ was assumed to be constant (i.e.  $V_{A0}^*=1$) in the models of McLaughlin \& Hood (\citeyear{MH2004}; \citeyear{MH2005}; etc) and there was no possibility of phase mixing, i.e. we have now removed a key assumption of these previous models.


\subsection{Method of Characteristics and D'Alembert solution}\label{sec:6.5.1.1}

Equation (\ref{equation_alfven_wave}) can be solved using the method of characteristics. Let $\frac {d }{d s} =  { \mathbf{B}}_0 \cdot \nabla =   x \frac {\partial }{\partial x} - z \frac {\partial }{\partial z } $, where $s$ is a parameter along a characteristic, and compare with, e.g.,  $\frac {d v_y}{d s} =   \frac {d x}{d s} \frac {\partial v_y}{\partial x} +  \frac {d z}{d s} \frac {\partial v_y}{\partial z }$. Comparing like terms yields:
\begin{eqnarray}
x = x_0 e^s \; , \quad z = z_0 e^{-s} \; \label{exponential_2}, 
\end{eqnarray}
where $x_0$ and $z_0$ are the starting positions of our characteristics. Thus,  $s= -\log { \frac{z}{z_0} } = \log { \frac{x}{x_0} } $. Thus equation (\ref{equation_alfven_wave}) can be written:
\begin{eqnarray}
\frac {\partial^2 v_y }{\partial t^2} =\frac{1}{\rho_0(xz)} \frac {d^2 }{d s^2}  v_y  = V_{A0}^2  \frac {d^2 }{d s^2}  v_y \; .
\end{eqnarray}
This characteristic equation can be solved with a D'Alembert solution such that:
\begin{eqnarray}
v_y = \mathcal{F} \left[ t - \sqrt{\rho_0} \:s  \right] + \mathcal{G} \left[ t + \sqrt{\rho_0} \:s  \right] =  \mathcal{F} \left[ t -   \frac {s} {V_{A0}{(A_0)}}  \right] + \mathcal{G} \left[ t + \frac {s} {V_{A0}{(A_0)}}  \right]\label{dalembert}
\end{eqnarray}
where $\mathcal{F}$ and  $\mathcal{G}$ are functions prescribed by the initial/boundary conditions. Note here that we can only implement the D'Alembert solution since $A_0$ is a constant along each fieldline (characteristic).


\subsection{Equilibrium density profiles}\label{sec:6.5.1.1.densityprofile}

\begin{figure}
\begin{center}
\includegraphics[width=5.8in]{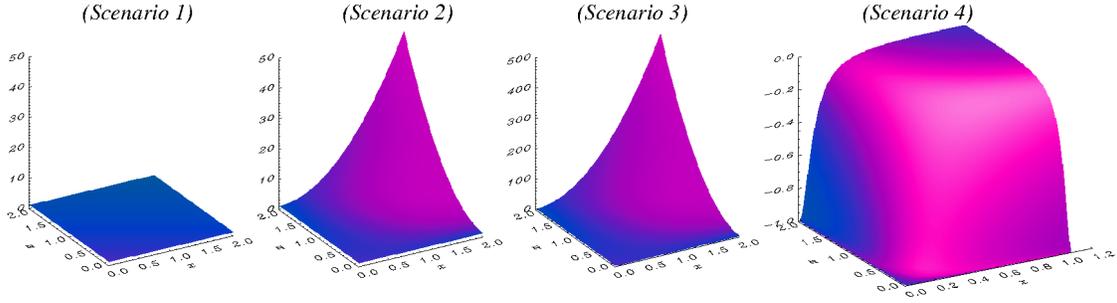}
\caption{ {\emph{Scenario 1}} shows density profile of scenario 1, i.e. $\rho_0=1$ (uniform density). {\emph{Scenario 2}} shows density profile of scenario 2, i.e. $\rho_0= 1+3 \left(xz\right)^2$ (weakly non-uniform density).  {\emph{Scenario 3}} shows density profile of scenario 3, i.e. $\rho_0= 1+30 \left(xz\right)^2$ (strongly non-uniform density, note change in $z-$axis).  {\emph{Scenario 4}} shows density profile of scenario 4, i.e.  $\rho_0={ \left[ {1+30 \left(xz\right)^2 } \right]} ^{-1}$. Since  $0 < \rho_0 \leq 1$ in scenario 4, this subfigure is presented as a surface of $-\rho_0$, as this shows the profile behaviour more clearly.}
\label{densityprofiles}
\end{center}
\end{figure}

In this paper, we are investigating the effect of including a non-uniform background density profile, and we present results from four scenarios. The first three cases will consider a  density profile  of the form $\rho_0 = 1+\lambda \left( xz \right)^2$, where we vary the parameter $\lambda$. Firstly, we will consider  a uniform density profile (where $\lambda =0$). This system is identical to that investigated in McLaughlin \& Hood (\citeyear{MH2004}),                   and provides an excellent visual comparison  to the other scenarios. Secondly, we consider a  weakly changing density profile of the form $\rho_0 = 1+3 \left(xz\right)^2$ ($\lambda=3$) and thirdly, we consider a more extreme density profile of the form $\rho_0 = 1+30 \left(xz\right)^2$ ($\lambda =30$). The second and third choices of density profile consider a region of highest $V_{A0}$, i.e. smallest $\rho_0$ {\emph{close to the null point}}. We will also consider a fourth scenario where $\rho_0 = \left[ {1+30\left( xz\right)^{2} } \right]^{-1}$, where the maximum $V_{A0}$ now occurs {\emph{away}}  from the null point. Tests show that these four  choices of density profile communicate all the general results well. These four density profiles can be seen in Figure \ref{densityprofiles} (note the axes vary between subfigures).

\begin{figure}
\includegraphics[width=5.8in]{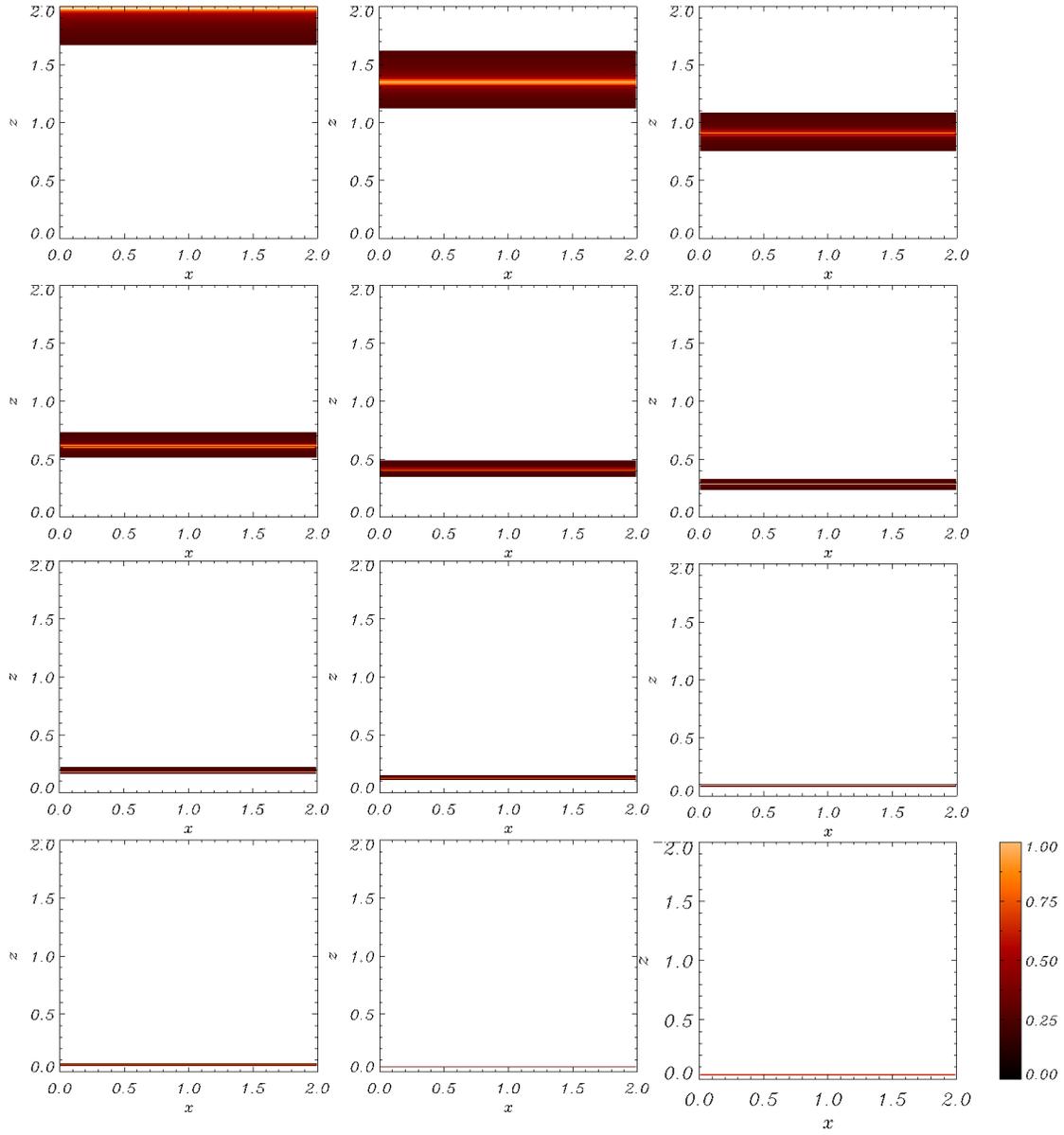}
\caption{Contours of $v_y$ for an Alfv\'en wave sent in from upper boundary for $0\leq x \leq 2$ and its resultant propagation at times $(a)$ $t$=0.25, $(b)$ $t$=0.6, $(c)$ $t$=1.0, $(d)$ $t=1.4$, $(e)$ $t$=1.8, $(f)$ $t$=2.2, $(g)$ $t=2.6$, $(h)$ $t$=3.0, $(i)$ $t$=3.4, $(j)$ $t=3.8$, $(k)$ $t$=4.2, $(l)$ $t$=4.6, labelling from top left to bottom right.}
\label{fig:6.4.1}
\end{figure}

\begin{figure}
\includegraphics[width=5.8in]{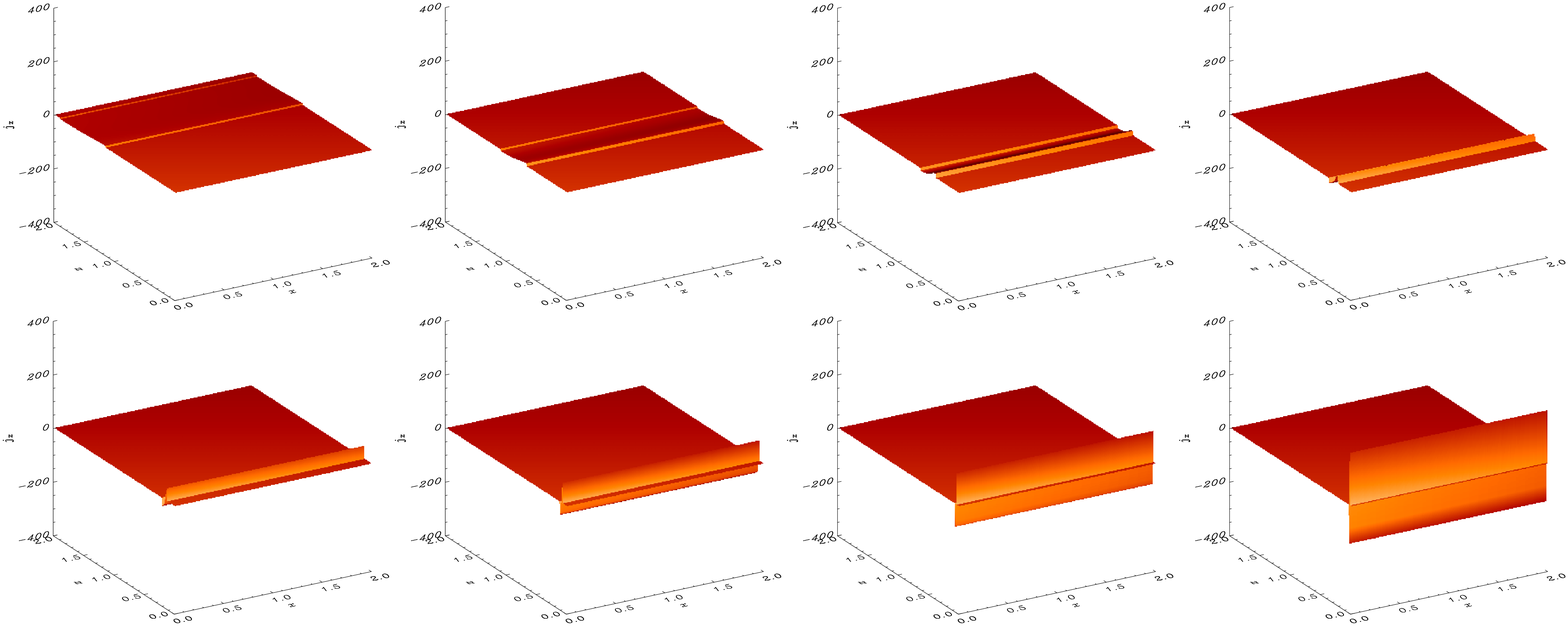}
\caption{Shaded surfaces showing the build-up of $j_x$ at times $(a)$ $t$=0.5, $(b)$ $t$=1.0, $(c)$ $t$=1.5, $(d)$ $t$=2.0, $(e)$ $t$=2.5, $(f)$ $t$=3.0, $(g)$ $t$=3.5 and $(h)$ $t$=4.0, labelling from top left to bottom right.}
\label{fig:6.4.1.1}
\end{figure}

\begin{figure}
\includegraphics[width=5.8in]{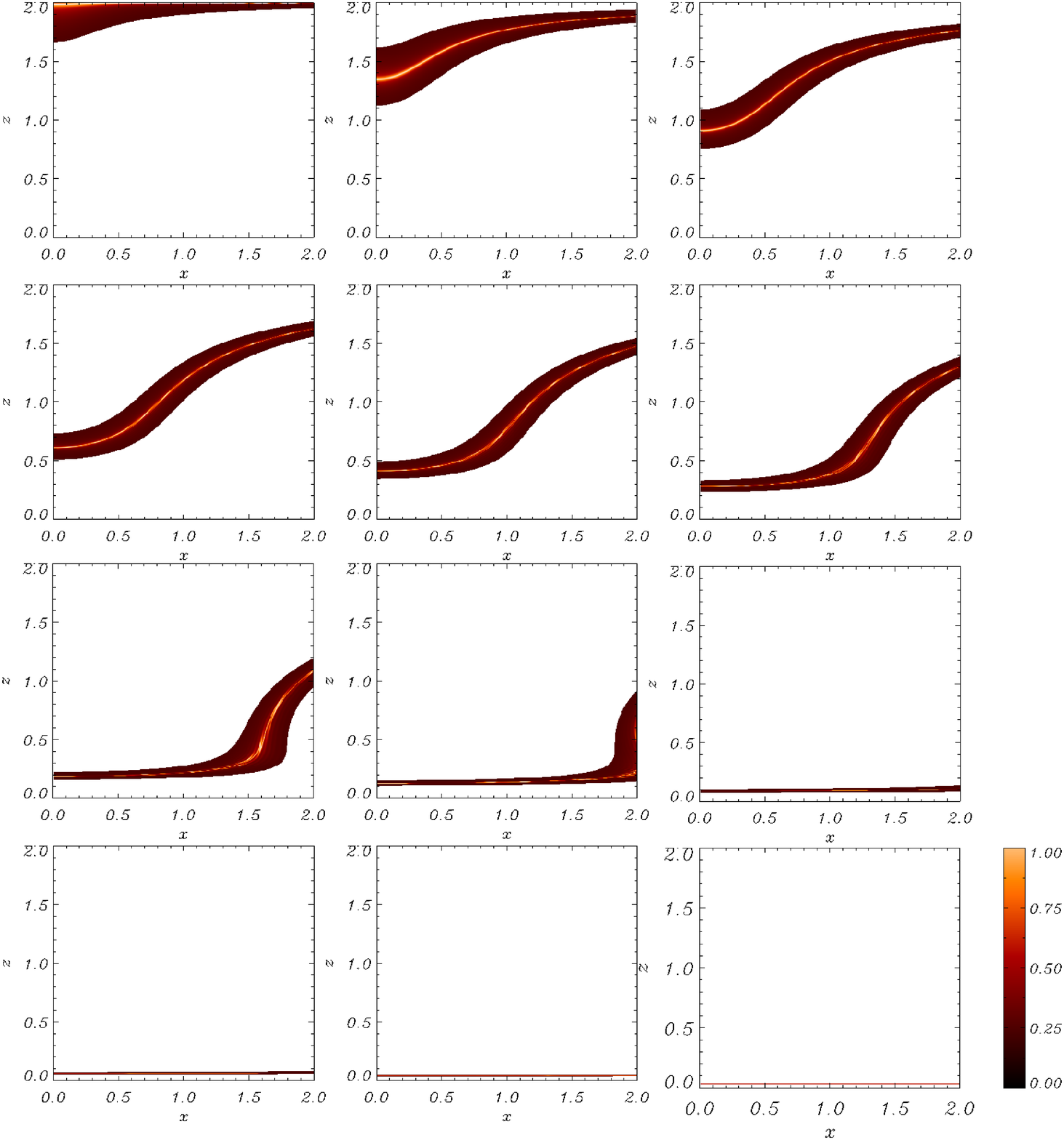}
\caption{Contours of $v_y$ for an Alfv\'en wave sent in from upper boundary for $0\leq x \leq 2$ and its resultant propagation at times $(a)$ $t$=0.25, $(b)$ $t$=0.6, $(c)$ $t$=1.0, $(d)$ $t=1.4$, $(e)$ $t$=1.8, $(f)$ $t$=2.2, $(g)$ $t=2.6$, $(h)$ $t$=3.0, $(i)$ $t$=3.4, $(j)$ $t=3.8$, $(k)$ $t$=4.2, $(l)$ $t$=4.6, labelling from top left to bottom right.}
\label{fig:6.5.1}
\end{figure}

\begin{figure}
\includegraphics[width=5.8in]{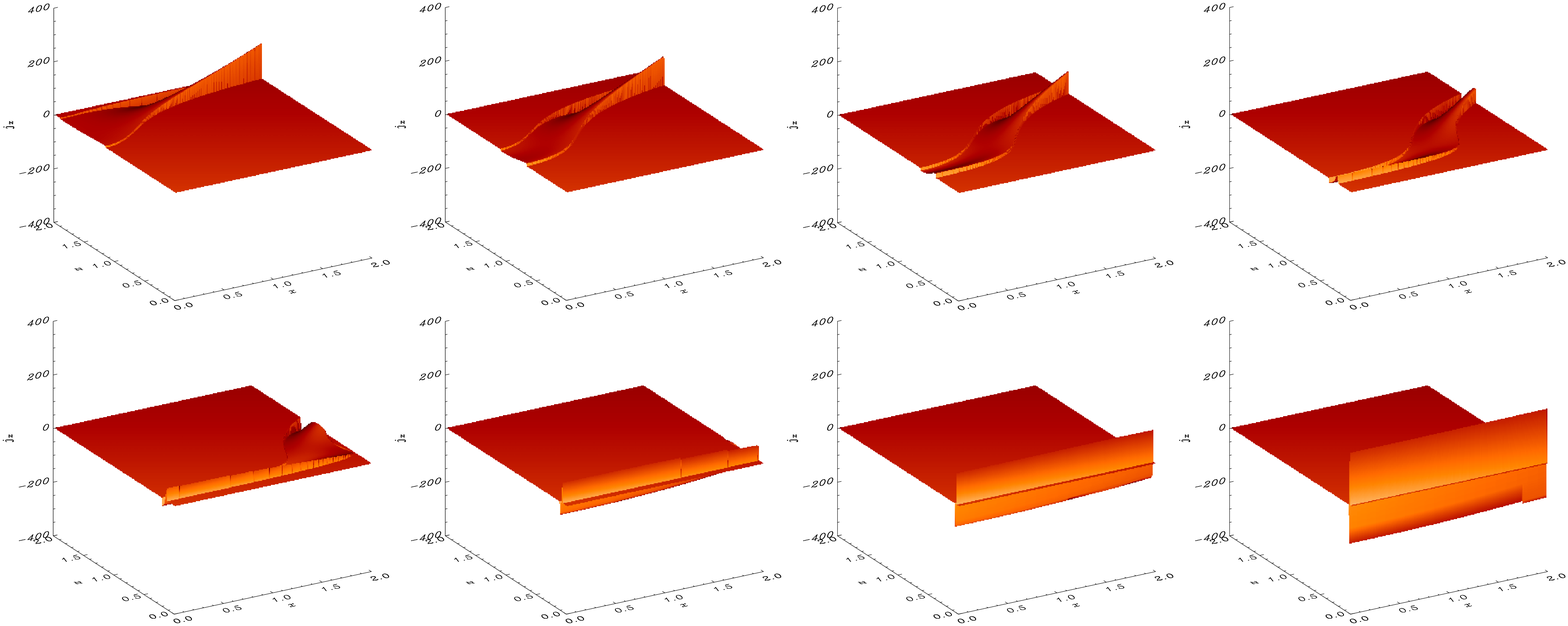}
\caption{Shaded surfaces showing the build-up of $j_x$ at times $(a)$ $t$=0.5, $(b)$ $t$=1.0, $(c)$ $t$=1.5, $(d)$ $t$=2.0, $(e)$ $t$=2.5, $(f)$ $t$=3.0, $(g)$ $t$=3.5 and $(h)$ $t$=4.0, labelling from top left to bottom right.}
\label{fig:6.5.1.2.1}
\end{figure}


\section{Numerical Simulations \& Analytical Solutions}\label{sec:6.4.1}

In this section, we solve equations (\ref{xmen})  numerically using a two-step Lax-Wendroff scheme and we present results from four non-uniform density scenarios. We drive our system with a wave pulse along the entire upper boundary, and we present a computational domain ($0 \leq x \leq 2$, $0 \leq z \leq 2$) with a single wave pulse coming in across the top boundary ($x=2$). The boundary conditions were set such that:
\begin{eqnarray}
\begin{array}{cl}
v_y(x, 2, t) = \sin { \left(  \omega t \right) } & {\mathrm{for} \; \; \left\{\begin{array}{c}  {0 \leq x \leq 2} \\ {0 \leq t \leq \frac {\pi}{\omega} } \end{array}\right. } \\
{{v_y} = 0 } & { \mathrm{otherwise} }
\end{array} \; ,  \nonumber\\
\left.\frac {\partial v_y } {\partial x } \right| _{x=2} =0 \; , \qquad \left.\frac {\partial v_y } {\partial x }  \right| _{x=0} = 0 \; , \qquad \left.\frac {\partial v_y } {\partial z }  \right| _{z=0}  = 0 \; .\label{PM_BC}
\end{eqnarray}
{{Results presented in this paper have a typical numerical resolution of $2000 \times 2000$ and (successful) convergence tests were performed.}} As detailed in $\S\ref{sec:6.5.1.1}$, the governing Alfv\'en wave equations can also be solved analytically (i.e. using equation \ref{dalembert}). In order to compare the analytical and numerical results, we must substitute the same initial conditions into the D'Alembert solution, i.e. $\mathcal{F} \left( t  \right) = \sin {\left( \omega t \right)}$ to get the analytical solution for $v_y$, namely:
\begin{eqnarray}
v_y (x,z,t) =\sin { \omega \left( t + \sqrt{\rho_0(xz)} \log { \frac{z}{z_0} } \right) } \quad \mathrm{for} \; \left\{ \begin{array}{c} {0 \leq t + \sqrt{\rho_0(xz)}}\log { \frac {z}{z_0} } \leq {\frac {\pi}{\omega}} \\ {0 \leq {x_0} \leq 2} \end{array} \right.  \; .\label{11-12}
\end{eqnarray}
It should also be noted that the agreement between all the numerical and analytical work in this paper is excellent. 

\vspace{0.05cm}

We can also use our  D'Alembert solution to calculate $b_y$ and hence $j_x$ and $j_z$. For the first three scenarios;  $\rho_0 = 1+\lambda \left(xz\right)^2$, and  so using equation (\ref{11-12}) we can write:
\begin{eqnarray}
b_y &=& -  \sqrt{1+\lambda x^2 z^2} \sin {\left[ \omega \left( t + \sqrt{1+\lambda x^2 z^2}\log {\frac {z}{z_0} } \right) \right]} \; ,                              \label{analytical_BY}         \\
j_x  &=& \left[ {\frac {\omega   \left(1+\lambda x^2 z^2\right) }{z}   }       +  {  \omega  {\lambda x^2 z}\log { \frac {z}{z_0} }  }    \right] \cos {\left[ \omega \left( t +  \sqrt{1+\lambda x^2 z^2}\log { \frac{z}{z_0} } \right) \right] } \nonumber \\
&&+ \frac {\lambda x^2 z}{\sqrt{ 1+\lambda x^2 z^2}}         \sin {\left[ \omega \left( t +  \sqrt{1+\lambda x^2 z^2}\log { \frac{z}{z_0} } \right) \right]}
  \label{analytical_JX}  \\
j_z  &=&   -\frac {\lambda x z^2}{\sqrt{ 1+\lambda x^2 z^2}}     \sin {\left[ \omega \left( t +  \sqrt{1+\lambda x^2 z^2}\log { \frac{z}{z_0} }\right)\right]} \nonumber  \\
&&- \omega \log { \left(\frac{z}{z_0}\right)} \lambda x z^2 \cos{\left[ \omega \left( t +  \sqrt{1+\lambda x^2 z^2}\log { \frac{z}{z_0}}\right)\right] }   \label{analytical_JZ} 
\end{eqnarray}
Note that the analytical solution for $b_y$, $j_x$ and $j_z$ is slightly different for  $\rho_0 = 1 / \left[  {1+30 \left(xz\right)^2} \right] $.


\subsection{Scenario 1 : Uniform Density ($\lambda=0$)}\label{sec:6.4}

The first three scenarios will consider a  density profile  of the form $\rho_0 = 1+\lambda \left( xz \right)^2$, where we vary the parameter $\lambda$. Firstly, we will consider  a uniform density profile (where $\lambda =0$). The resultant wave evolution can be seen in Figure \ref{fig:6.4.1}.

\vspace{0.05cm}

We find that the linear Alfv\'en wave propagates downwards from the top boundary and begins to spread out, following the fieldlines. As the wave approaches the $x-$axis (the separatrix), it thins but keeps its original amplitude. The wave eventually accumulates very near the separatrix. Note that these results are similar  to those investigated in McLaughlin \& Hood (\citeyear{MH2004}), and are presented here to provide a visual comparison and contrast to the other scenarios (McLaughlin \& Hood \citeyear{MH2004} actually used a different driver, making direct comparisons with scenarios 2, 3 \& 4 less obvious).

\vspace{0.05cm}

We can also solve equation (\ref{alfvenalpha_PM}) using our D'Alembert solution. Here, equations (\ref{analytical_BY} - \ref{analytical_JZ}) can be simplified under $\lambda=0$ to give:
\begin{eqnarray*}
b_y = -  \sin { \omega \left( t + \log { \frac {z}{z_0} } \right) } \; \;, \;\; j_x  ={\frac {\omega  }{z}   }     \cos { \omega \left( t + \log { \frac{z}{z_0} } \right) }   \;\; , \;\; j_z  =  0   \;\;  .
\end{eqnarray*}
Hence, the Alfv\'en wave causes current density to build up along the separatrix. Furthermore, since $z = z_0 e^{-s}$ from equation (\ref{exponential_2}), we see that this $j_x$ build up is {\emph{exponential}} in time (due to the $1/z$ dependence) whereas  $j_z=0$ for all time. Figure \ref{fig:6.4.1.1} shows the build-up of $j_x$.



\subsection{Scenario 2 : Weakly non-uniform Density  ($\lambda=3$)  }\label{sec:6.5.1}

We now consider a weakly non-uniform density profile: $\rho_0(x,z) = {1+3 x^2z^2}$. As in scenario 1 in $\S\ref{sec:6.4}$,  equations (\ref{xmen})  are  solved numerically using our two-step Lax-Wendroff scheme, utilising the same boundary and initial conditions (equations \ref{PM_BC}) but now implementing our weakly-changing density profile, i.e. the governing Alfv\'en  wave equation is now:
\begin{eqnarray*}
  \frac {\partial^2 v_y }{\partial t^2} =\frac{1}{1+3 x^2z^2}  \left(x \frac {\partial }{\partial x} - z\frac {\partial }{\partial z} \right) ^2 v_y \; \label{alfvenalpha_PM_3}.
\end{eqnarray*}

\vspace{0.05cm}

The results for $v_y$ can be seen in Figure \ref{fig:6.5.1}. We see that the Alfv\'en wave again descends from the upper boundary ($x=2$) and accumulates along the separatrix ($x-$axis), but now the (initially  planar) wave is distorted; a phenomenon not seen in previous null point studies. The varying speed, $V_{A0}(xz)$, means different fluid elements of the wave travel at different speeds. Thus, the fluid elements of the wave closest to the $x=0$ axis, where $V_{A0}(xz)$ takes its maximum value (or alternatively $\rho_0(xz)$ takes its minimum value) propagate at a greater speed than those fluid elements away from the axis (i.e. left-hand-side propagates faster than right). Thus, the wave is distorted and descends at different rates. This is clearly significantly different wave behaviour to that of the uniform density case considered in Figure  \ref{fig:6.4.1}. The wave does however still eventually accumulate  along the separatrix.

\vspace{0.05cm}

\begin{figure}[t]
\includegraphics[width=5.8in]{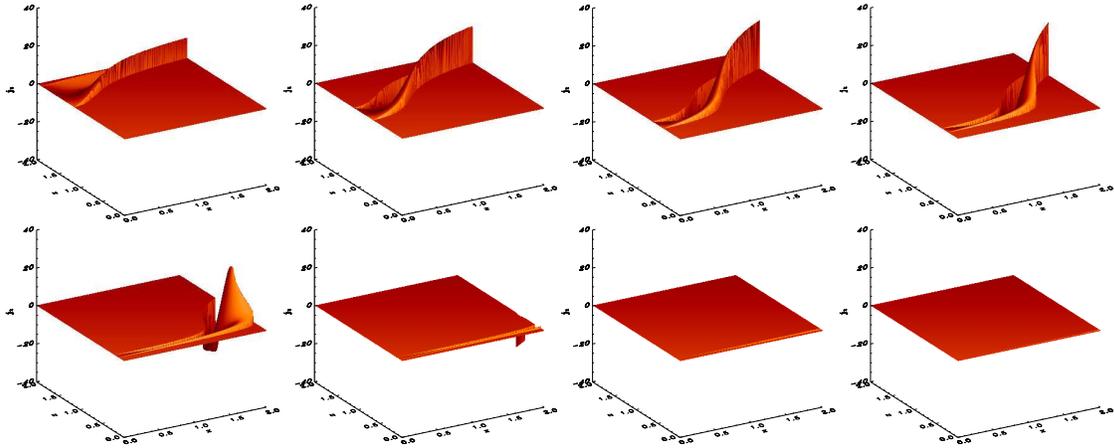}
\caption{Shaded surfaces showing the build-up of $j_z$ at times $(a)$ $t$=0.5, $(b)$ $t$=1.0, $(c)$ $t$=1.5, $(d)$ $t$=2.0, $(e)$ $t$=2.5, $(f)$ $t$=3.0, $(g)$ $t$=3.5 and $(h)$ $t$=4.0, labelling from top left to bottom right.}
\label{fig:6.5.1.2.2}
\end{figure}

Again, the D'Alembert solution agrees exactly with the numerical simulation. Substituting  $\lambda=3$ into equations (\ref{analytical_JX}) and (\ref{analytical_JZ}) gives  analytical forms for $j_x$ and $j_z$. These can be seen in Figures \ref{fig:6.5.1.2.1} and \ref{fig:6.5.1.2.2}, respectively. In Figure \ref{fig:6.5.1.2.1}, we can see that there is a large concentration of $j_x$ initially along the wave, due to the changing density profile. This concentration propagates with the wave and begins to accumulate along the separatrix ($x-$axis). The build-up of $j_x$ along the separatrix is substantially more than in early subfigures. In Figure  \ref{fig:6.5.1.2.2}, we can see that there is initially a concentration of $j_z$ (due to the changing density profile) but that this decays away as time elapses. Hence  the separatrix is still be the location for preferential heating due to  Alfv\'en waves, even with the inclusion of a (weakly) non-uniform density profile.


\subsection{Scenario 3 : Strongly  non-uniform Density ($\lambda=30$)  }\label{sec:6.5.2}

\begin{figure}
\includegraphics[width=5.8in]{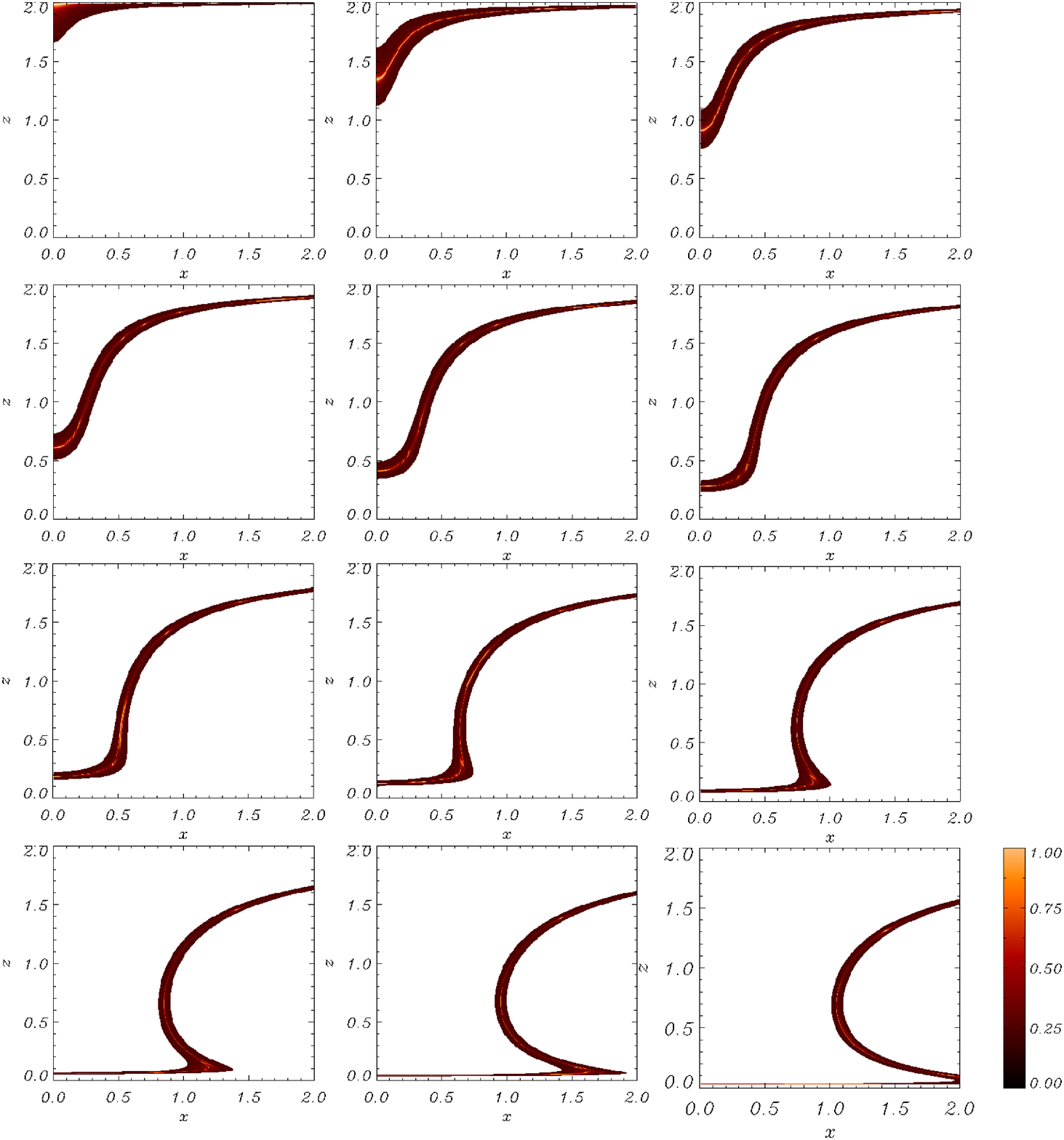}
\caption{Contours of $v_y$ for an Alfv\'en wave sent in from upper boundary for $0\leq x \leq 2$ and its resultant propagation at times $(a)$ $t$=0.25, $(b)$ $t$=0.6, $(c)$ $t$=1.0, $(d)$ $t=1.4$, $(e)$ $t$=1.8, $(f)$ $t$=2.2, $(g)$ $t=2.6$, $(h)$ $t$=3.0, $(i)$ $t$=3.4, $(j)$ $t=3.8$, $(k)$ $t$=4.2, $(l)$ $t$=4.6, labelling from top left to bottom right.}
\label{fig:6.5.2}
\end{figure}

\begin{figure}[t]
\includegraphics[width=5.8in]{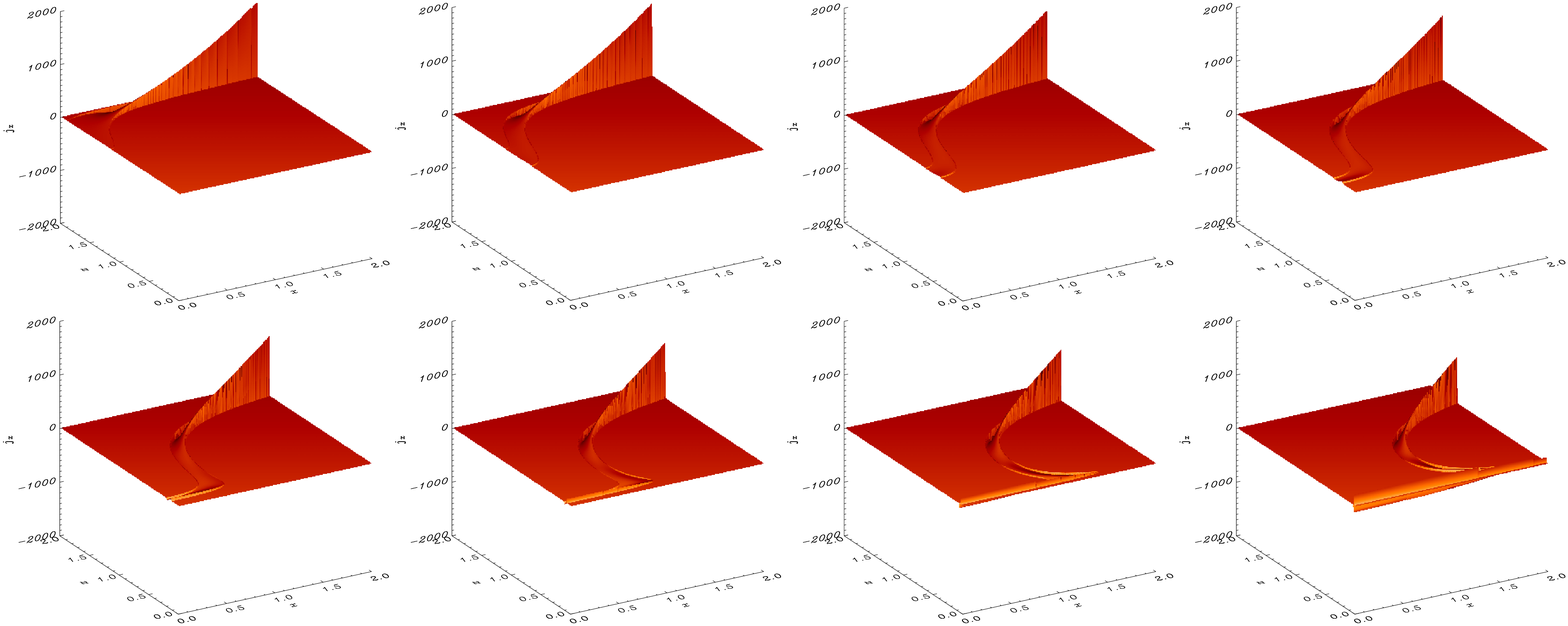}
\caption{Shaded surfaces showing the build-up of $j_x$ at times $(a)$ $t$=0.5, $(b)$ $t$=1.0, $(c)$ $t$=1.5, $(d)$ $t$=2.0, $(e)$ $t$=2.5, $(f)$ $t$=3.0, $(g)$ $t$=3.5 and $(h)$ $t$=4.0, labelling from top left to bottom right.}
\label{fig:6.5.2.2.1}
\end{figure}

We now consider a strongly non-uniform density profile: $\rho_0(x,z) = {1+30 x^2z^2}$. As in scenarios 1 \& 2 above,  equations (\ref{xmen})  are  solved numerically using a two-step Lax-Wendroff scheme, utilising the same boundary and initial conditions (equations \ref{PM_BC}) but now implementing our strongly-changing density profile, i.e. the governing Alfv\'en  wave equation is now:
\begin{eqnarray*}
  \frac {\partial^2 v_y }{\partial t^2} =\frac{1}{1+30 x^2z^2}  \left(x \frac {\partial }{\partial x} - z\frac {\partial }{\partial z} \right) ^2 v_y \; \label{alfvenalpha_PM_4}.
\end{eqnarray*}

\vspace{0.05cm}

The resultant propagation of $v_y$ can be seen in Figure \ref{fig:6.5.2}. The Alfv\'en  wave behaviour is similar to that seen for the weakly-changing density profile ($\S\ref{sec:6.5.1}$), but with one important distinction. Again, we see that the Alfv\'en wave descends and accumulates along the separatrix. The wave is distorted from its original planar form by the varying density profile, and hence different parts of the wave descend at different speeds. Thus, the wave travels faster nearer to the $z-$axis than away from it. However, since the Alfv\'en wave is confined to the fieldlines, and propagating along those fieldlines, there comes a point where fluid elements of the wave are so ahead of other elements of the wave that the wavefront (made by joining up all the elements at the same $s$ value) actually bends back upon itself; again a phenomenon not seen before in null point investigations. This can be seen most clearly in the lower subfigures of  Figure \ref{fig:6.5.2}. Despite this however, once again the Alfv\'en wave still eventually accumulate  along the separatrix ($x-$axis).

\vspace{0.05cm}

Since the wave is so stretched  where it forms the \lq{reflection point}\rq{}  in the wavefront, there may be a great deal of current build-up near this point. Hence, this may provide an additional location for (preferential) heating, and so we investigate the resultant $j_x$ and $j_z$.

\vspace{0.05cm}

\begin{figure}[t] 
\includegraphics[width=5.8in]{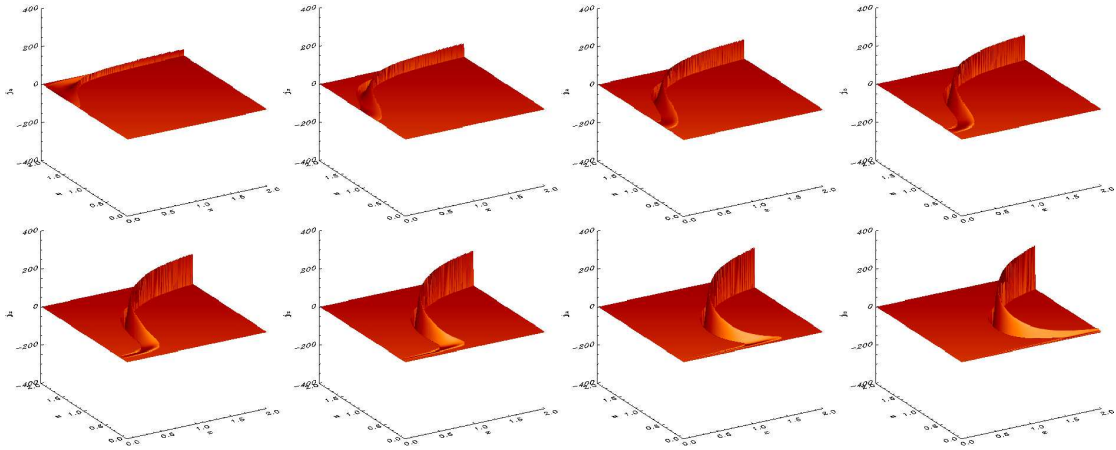}
\caption{Shaded surfaces showing the build-up of $j_z$ at times $(a)$ $t$=0.5, $(b)$ $t$=1.0, $(c)$ $t$=1.5, $(d)$ $t$=2.0, $(e)$ $t$=2.5, $(f)$ $t$=3.0, $(g)$ $t$=3.5 and $(h)$ $t$=4.0, labelling from top left to bottom right.}
\label{fig:6.5.2.2.2}
\end{figure}

As before, the D'Alembert solution agrees exactly with the numerical simulation. Substituting  $\lambda=30$ into equations (\ref{analytical_JX}) and (\ref{analytical_JZ}) gives  analytical forms for $j_x$ and $j_z$, and these can be seen in Figures \ref{fig:6.5.2.2.1} and \ref{fig:6.5.2.2.2}. In Figure \ref{fig:6.5.2.2.1},  we can see that there is indeed a very large concentration (note value on axis!) of $j_x$ initially along the wave, due to the changing density profile. This current concentration propagates co-spatially with the wave and begins to accumulate along the separatrix ($x-$axis).  This build-up along the separatrix eventually overtakes  the magnitude of current concentrations elsewhere (although this occurs at a later time than that shown in the  last subfigure). Hence, the separatrix will {\emph{still}} be the location for the majority of heating, however small $\eta$ is taken to be. Of course, if  $\eta$ is taken to be extremely large (unphysical) then there may be some heating along other parts of the Alfv\'en   wave. However, the density profile  invoked in this scenario has a very  extreme distribution and so perhaps under coronal conditions this scenario would not take place and, consequently,  preferential heating would {\emph{still occur along the separatrices}}.

\vspace{0.05cm}

In Figure  \ref{fig:6.5.2.2.2}, we can see that there is initially a concentration of $j_z$ (due to the changing density profile). This  propagates in the increasing $x$ and decreasing $z$ directions, and decays away near the $x-$axis. The rest of the $j_z$ concentration propagates away and out of our box, but will eventually decay away (like $j_z$ did before). This can be see in  the form of $j_z$ from equations (\ref{analytical_JZ}).

\vspace{0.05cm}

It is perhaps not clear from Figures \ref{fig:6.5.2.2.1} and \ref{fig:6.5.2.2.2}  alone that the maximum current  build-up (still)  occurs at the $x-$axis ($z=0$ line). To show this analytically, we can utilise equations (\ref{analytical_JX}) and (\ref{analytical_JZ}) in combination with  equations (\ref{exponential_2}). Thus, we can  substitute $x = x_0 e^s$, $z = z_0 e^{-s}$ and hence  $xz=x_0z_0$  to give:
\begin{eqnarray}
j_x  &=& e^s  \left\{  \left[ {\frac {\omega   \rho_0 }{z_0}   }      - { \left(\omega  {\lambda x_0^2z_0} \right)  s  }    \right] \cos { \omega \left( t - \sqrt{\rho_0}s\right) } + \frac {\lambda x_0^2 z_0}{\sqrt{\rho_0}}         \sin { \omega \left( t -  \sqrt{\rho_0}s \right) }\right\}     \nonumber  \\
 &=& e^s  \left[  \left( B     - C \:s      \right) \cos { \omega \left( t - \sqrt{\rho_0}s\right) } +   D      \sin { \omega \left( t -  \sqrt{\rho_0}s \right) }\right]
   \nonumber      \\
j_z  &=& e^{-s} \left[  -\frac {\lambda x_0 z_0^2}{\sqrt{ \rho_0}}     \sin { \omega \left( t -  \sqrt{\rho_0}s\right)} + \left(\omega  \lambda x_0 z_0^2 \right) s \cos{ \omega \left( t -  \sqrt{\rho_0}s  \right)   }    \right] \nonumber \\
 &=& e^{-s} \left[  -  E   \sin { \omega \left( t -  \sqrt{\rho_0}s\right)} +  F \:s  \cos{ \omega \left( t -  \sqrt{\rho_0}s  \right)   }    \right]\label{turk}
\end{eqnarray}
where $x_0$, $z_0$, $\omega$, $\lambda$, $\rho_0$, $B$, $C$, $D$, $E$ and $F$ are all constants for a specific fieldline ( $B$,...,$F$  are just collected constants but are all strictly positive).

\vspace{0.05cm}

\begin{figure}
\begin{center}
\includegraphics[width=3.25in]{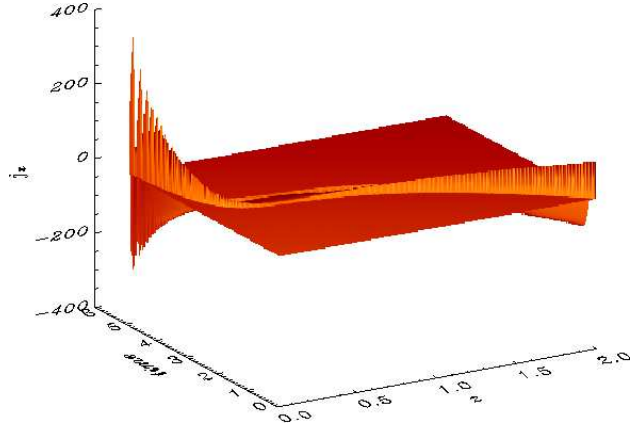}
\caption{Shaded surfaces showing the build-up of $j_x$ plotted against $z$ and time with $x=0.5$.}
\label{buildupzandtime}
\end{center}
\end{figure}

Figure \ref{buildupzandtime} shows a surface of the build-up of $j_x$ plotted against $z$ and time, along $x=0.5$. Here, we can see that $z$ decreases from $z=2$ down to near $z=0$ as time elapses and that $j_x$ is building up the closer we are to $z=0$ and the build-up is increasing in time.

\vspace{0.05cm}

Hence, we can see now clearly see that the behaviour of $j_x$ and $j_z$ follow complicated forms that depend upon many starting parameters and on $s$, but that at large times, $j_x$ will eventually build-up exponentially and  $j_z$ will decay exponentially. Thus, for the linear Alfv\'en wave,  preferential heating will still occur along the separatrices, despite the inclusion of either weakly or strongly  non-uniform density profiles.

\subsection{Scenario 4 : Non-uniform Density  $\rho_0 = \left[{1+30\left( xz\right)}^2\right]^{-1}$ }\label{sec:6.5.5}

\begin{figure}
\includegraphics[width=5.8in]{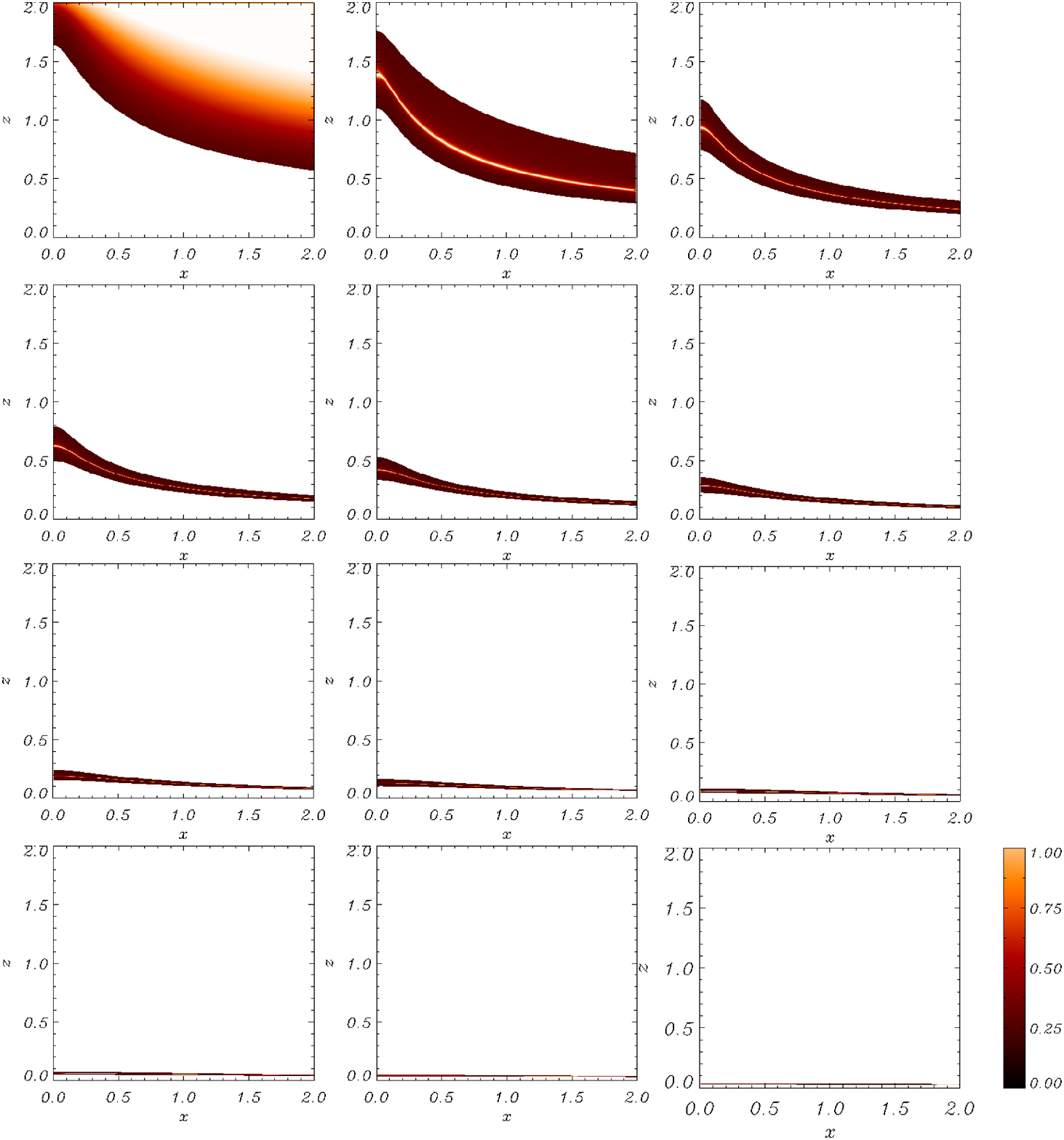}
\caption{Contours of $v_y$ for an Alfv\'en wave sent in from upper boundary for $0\leq x \leq 2$ and its resultant propagation at times $(a)$ $t$=0.25, $(b)$ $t$=0.6, $(c)$ $t$=1.0, $(d)$ $t=1.4$, $(e)$ $t$=1.8, $(f)$ $t$=2.2, $(g)$ $t=2.6$, $(h)$ $t$=3.0, $(i)$ $t$=3.4, $(j)$ $t=3.8$, $(k)$ $t$=4.2, $(l)$ $t$=4.6, labelling from top left to bottom right.}
\label{fig:6.5.5.1.final}
\end{figure}

\begin{figure}
\includegraphics[width=5.8in]{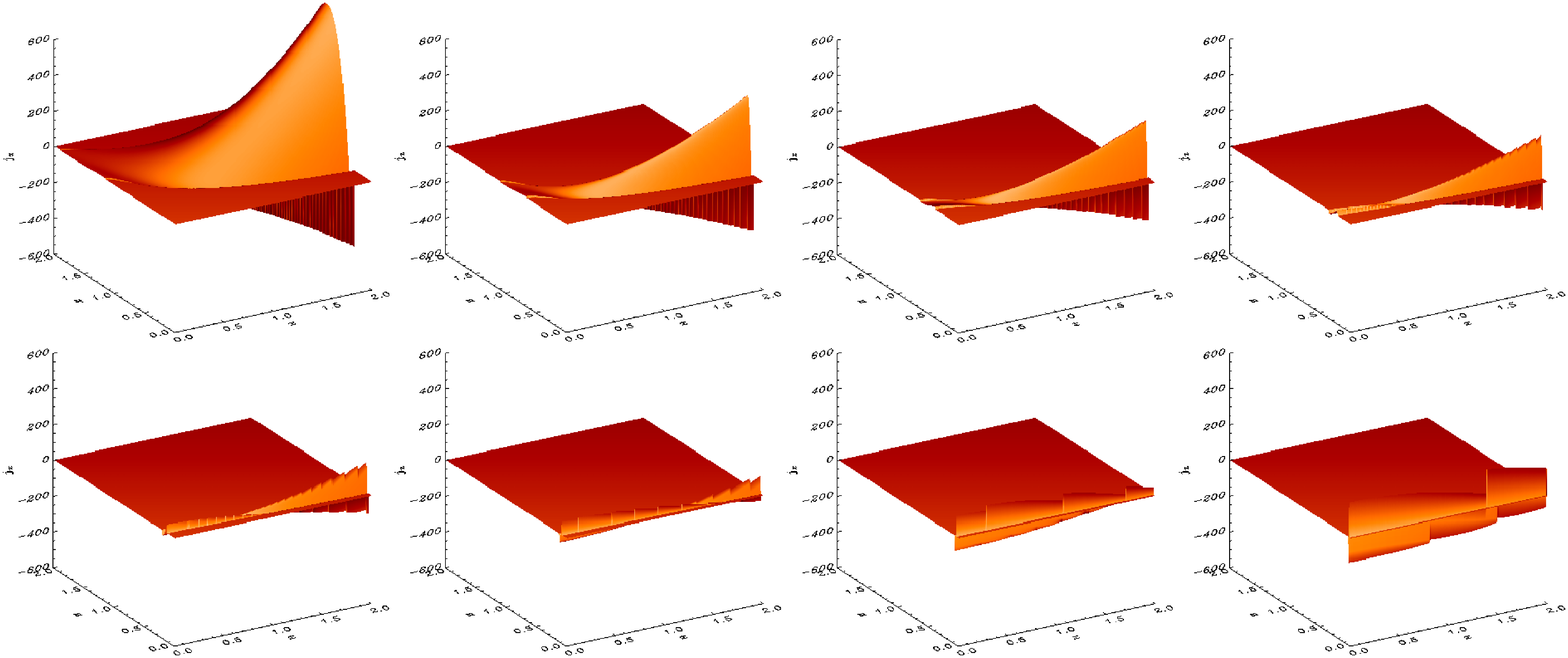}
\caption{Shaded surfaces showing the build up of $j_x$ at times $(a)$ $t$=0.5, $(b)$ $t$=1.0, $(c)$ $t$=1.5, $(d)$ $t$=2.0, $(e)$ $t$=2.5, $(f)$ $t$=3.0, $(g)$ $t$=3.5 and $(h)$ $t$=4.0, labelling from top left to bottom right.}
\label{fig:6.5.5.2}
\end{figure}

We now consider our final  non-uniform density profile: $\rho_0 = \left[{1+30\left( xz\right)}^2\right]^{-1}$. As before, equations (\ref{xmen})  are  solved numerically using a two-step Lax-Wendroff scheme, with the same boundary and initial conditions  (given by  equations \ref{PM_BC}) but now implementing our fourth  non-uniform density profile. Thus, the governing  Alfv\'en wave equation is now:
\begin{eqnarray*}
  \frac {\partial^2 v_y }{\partial t^2} =\left( {1+30 x^2z^2} \right) \left(x \frac {\partial }{\partial x} - z\frac {\partial }{\partial z} \right) ^2 v_y \; \label{alfvenalpha_PM_7}.
\end{eqnarray*}

\vspace{0.05cm}

The results for $v_y$ can be seen in Figure \ref{fig:6.5.5.1.final}. Here, the greatest magnitude of  $V_{A0}$ occurs away from the axes (in these numerical boxes the maximum occurs at $x=z=2$). Thus, we find that the Alfv\'en wave is again distorted from its initially planar shape, but that it now travels faster the {\emph{further}} we are away from the null point / axes. The first subfigure of  Figure \ref{fig:6.5.5.1.final} shows the massive speed differential across the wave after a very short time. The linear Alfv\'en wave then descends and starts to accumulate along the $x-$axis (separatrix). Here, the wave slows down and thins, but keeps its original amplitude. There is no spike / reflection point formed, as there was in $\S\ref{sec:6.5.2}$. This is clearly a different velocity profile to that seen in the previous three figures (i.e. Figs \ref{fig:6.4.1}, \ref{fig:6.5.1} \& \ref{fig:6.5.2.2.1}) but the phenomenon of different fluid elements propagating  at different speeds due to the non-uniform density profile is common to all scenarios.

\vspace{0.05cm}

\begin{figure}[t]
\includegraphics[width=5.8in]{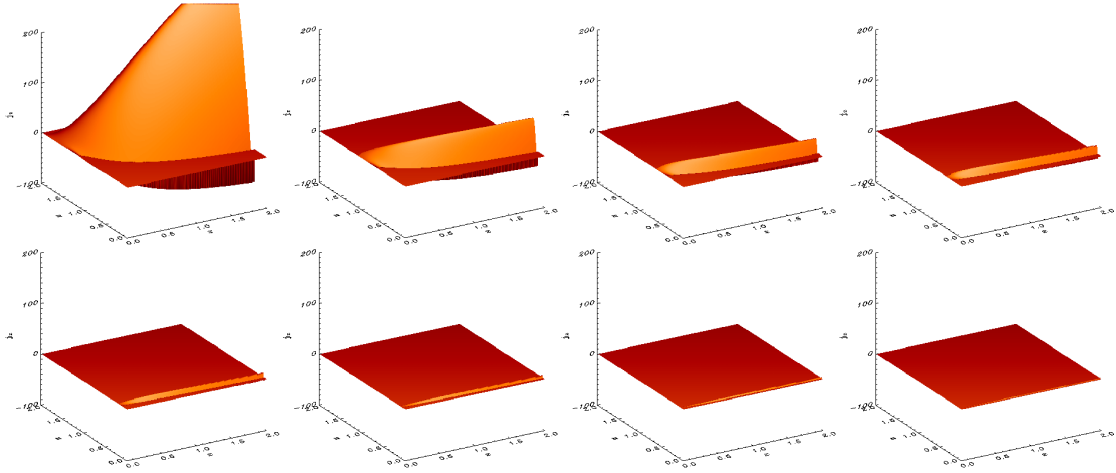}
\caption{Shaded surfaces showing the build up of $j_z$ at times $(a)$ $t$=0.5, $(b)$ $t$=1.0, $(c)$ $t$=1.5, $(d)$ $t$=2.0, $(e)$ $t$=2.5, $(f)$ $t$=3.0, $(g)$ $t$=3.5 and $(h)$ $t$=4.0, labelling from top left to bottom right.}
\label{fig:6.5.5.3}
\end{figure}

As before, the D'Alembert solution agrees exactly with the numerical simulation, and we can use our  D'Alembert solution to work out $b_y$,  $j_x$ and $j_z$ (as we did in  equations \ref{analytical_BY}-\ref{analytical_JZ}) for this fourth density profile. The resultant behaviour for $j_x$ and $j_z$ can be seen in Figures \ref{fig:6.5.5.2} and \ref{fig:6.5.5.3}. In  Figure \ref{fig:6.5.5.2}, we can see that there is initially a large concentration of $j_x$ due to the extreme density profile, but that this then decays away (as the wave propagates along the fieldlines and out of the box). At a later time, after the wave is near the separatrix, $j_x$ starts to grow again. In Figure \ref{fig:6.5.5.3},  we can see that there is initially a large concentration of $j_z$ (due to the changing density profile), but that it very quickly decays away. Thus, there is a large current accumulation  along the separatrix and our key result about preferential Alfv\'en wave heating again  holds.

\vspace{0.05cm}

As mentioned above, our  D'Alembert solution gives us general forms for $b_y$,  $j_x$ and $j_z$.  Substituting  $\rho_0 = \left[{1+\gamma\left( xz\right)}^2\right]^{-1}$ into these forms (where $\gamma=30$ in this study case but $\gamma$ is used so the result is more general)  and  substituting $xz=x_0z_0$ and $x = x_0 e^s$, $z = z_0 e^{-s}$ gives:
\begin{eqnarray*}
j_x  &=& e^s \left[ \left({ \omega \gamma x_0^2 z_0^2}{\rho_0} \: s  + \frac{\omega}{z_0} \right) \cos { \omega \left( t - \sqrt{\rho_0}s\right) } - \rho_0^{\frac{3}{2}} \gamma \:x_0^2z_0   \sin { \omega \left( t -  \sqrt{\rho_0}s\right)}         \right]\\
&=& e^s \left[ \left( B \: s + C   \right) \cos { \omega \left( t - \sqrt{\rho_0}s\right) } - D  \sin { \omega \left( t -  \sqrt{\rho_0}s\right)}         \right]\\
j_z  &=& e^{-s} \left[ -s \omega  \gamma x_0z_0^2 \rho_0^2 \cos { \omega \left( t - \sqrt{\rho_0}s\right) } + \gamma x_0z_0^2  \rho_0^{\frac{3}{2}}  \sin { \omega \left( t -  \sqrt{\rho_0}s\right)}    \right]\\
 &=& e^{-s} \left[ E \sin { \omega \left( t -  \sqrt{\rho_0}s\right)}   +  F\:s\cos { \omega \left( t - \sqrt{\rho_0}s\right) } \right]
\end{eqnarray*}
where again $x_0$, $z_0$, $\omega$, $\gamma$, $\rho_0$, $B$, $C$, $D$, $E$ and $F$ are all constants for a specific fieldline ($B$,...,$F$  are just collected constants but are all strictly positive). Note that these equations have a similar form to equations (\ref{turk}) but the constants $B$-$F$ are different.

\vspace{0.05cm}

Hence, we can see that the behaviour of $j_x$ and $j_z$ follow  complicated forms that depend upon many starting parameters and on $s$, but that (as before) $j_x$ will eventually build-up exponentially and  $j_z$ will decay exponentially. Thus, {{for the linear Alfv\'en wave,  preferential heating will still occur along the separatrices, even when a non-uniform density profile is considered}}.


\section{Conclusion}\label{sec:6.10}

We have  investigated the behaviour of the linear Alfv\'en wave in the neighbourhood of a 2D X-point geometry, investigating both uniform and non-uniform equilibrium density plasma. Specifically, we have considered four scenarios:
\begin{itemize}
\item{Uniform Density: $\rho_0=$constant}
\item{Weakly non-uniform Density: $\rho_0={1+3 x^2z^2}$ }
\item{Strongly non-uniform Density: $\rho_0={1+30 x^2z^2}$ }
\item{Non-uniform Density: $\rho_0 = \left[{1+30\left( xz\right)}^2\right]^{-1}$}
\end{itemize}



\vspace{0.05cm}

We find that the linear Alfv\'en wave  propagates along the equilibrium fieldlines and a single wave-fluid element is confined to the fieldline that it starts on. Since the wave strictly follows the fieldlines, it spreads out as it approaches the diverging null point. When a uniform plasma     density is considered, it was seen that the (initially planar) Alfv\'en wave front     remains purely planar, despite the varying equilibrium Alfv\'en-speed profile, and  that   the current density accumulates {\emph{exponentially}}    at the separatrices. In the non-ideal case, these Alfv\'enic disturbances   will  dissipate    their (wave) energy at these preferential locations.


\vspace{0.05cm}

We also investigated a variety of  non-uniform  equilibrium    density profiles, and found that in these scenarios  the (initially planar) wave front is now rapidly    stretched and distorted. In fact,  the wavefront is distorted from its initially planar shape and travels along the fieldlines at different speeds. Depending upon the exact form of the density profile, the wavefront can stretch so much that it bends back upon itself and creates a spike or \lq{reflection point}\rq{} - an effect not reported before in the case of wave behaviour around null points. However, in some cases, the creation of such a reflection point can require an extreme and unphysical density profile.

\vspace{0.05cm}

This paper set out to answer a key question: with the addition of a non-uniform density, and thus removing one of the key restrictions of McLaughlin \& Hood (\citeyear{MH2004}) and subsequent papers,  does the current density accumulation still occur preferentially at the separatrix or does phase mixing now allow the wave energy to be extracted from a different location? At its heart, the results in this paper  have  been all about the battle between {\emph{dissipation due to phase mixing}} and {\emph{dissipation of the current build-up along the separatrices}}. From our results  above, we conclude that the  current density build-up is limited except near     the separatrices. Thus, our  key result is that for the linear Alfv\'en wave preferential heating occurs along the separatrices, {\emph{even when a non-uniform density profile is considered}}.

\vspace{0.05cm}

{{

The energy carried by Alfv\'en waves is considered to play an important role in the heating of coronal holes and the acceleration of the solar wind (e.g. Ofman \& Davila  \citeyear{Ofman1995}; \citeyear{Ofman1997};  Chmielewski  {\emph{et al.}} {\citeyear{Chmielewski2013}}, and references therein). Our results highlight that the separatrices will be preferential locations for  Alfv\'en wave heating and thus present  a clear observational prediction.

}}

{{

Finally, this investigation has utilised linearised MHD equations (\S\ref{sec:2.2}) and this approach is only  valid when the perturbations in our physical parameters are much smaller than their equilibrium values. In addition, the  Alfv\'en wave is slowing down as it approaches the separatrices, hence its gradients are increasing, and moreover these have been shown to grow exponentially (see \S\ref{sec:6.4}, \ref{sec:6.5.1}, \ref{sec:6.5.2} \&  \ref{sec:6.5.5}). Thus, in a simple manner, our linearisation will start to break down on timescales  $t \simeq - \log {M_{\rm A}} $, where $M_{\rm A}$ is the initial Alfvén Mach number. Thus, our results are valid for small-to-medium amplitude  Alfv\'en waves, but further studies are required to fully understand the implications for large amplitude non-linear  Alfv\'en waves.

}}

\bigskip
{\bf Acknowledgements}

{The author acknowledges IDL support provided by STFC. JM  wishes to thank Alan Hood for insightful discussions and constant encouragement.}


\end{document}